# Structural stability, electronic band structure, and optoelectronic properties of quaternary chalcogenide $CuZn_2MS_4$ (M =In and Ga) compounds via first principles

**Anima Ghosh[1*] and R.Thangavel[2]**

[1]Department of Physics, School of Sciences and Humanities, SR University, Warangal, Telangana, India
[2]Department of Physics, Indian Institute of Technology (Indian School of Mines), Dhanbad, Jharkhand, India
*anima.ghosh@sru.edu.in; animaghosh10@gmail.com

**Abstract**

Quaternary chalcogenide compositions have been broadly explored due to their promising potential for various optoelectronic applications. The band structure, density of states and optical properties of $CuZn_2InS_4$ and $CuZn_2GaS_4$ for kesterite and stannite structures were studied with full potential augmented plane wave method (FP-LAPW) via Wien2k code. The total energy in equilibrium was calculated for different possible crystal structures and their phase stability, and transitions with *p-d* orbitals were analyzed. The absorption coefficient, dielectric function, and refractive index of these materials were also explored within a broad range of energy. We compared the calculated band gap values with available experimental results.

1. Introduction

In the studies of photovoltaic technologies, thermoelectric, and optoelectronic devices, the quaternary semiconductors $Cu_2$-II-IV-$VI_4$ have been paid tremendous attention since their elements are earth-abundance, inexpensive, and environment-friendly in nature. In addition, their band gap offers a broad range of the optical spectrum. The designs of $Cu_2$-II-III-$VI_4$ quaternary compositions are resulting from cation cross-substitution of binary II-VI and ternary I-III-$VI_2$. It offers an alternative route for exploring promising materials to use as absorber materials in photovoltaic applications. Originally, Goodman (1958) and Pamplin (1964) initiated the formation of quaternary chalcogenides from ternary I-III-$VI_2$ systems through cation cross-substitution which has obeyed octet rule [1-2]. There are two possibilities for these cation substitutions: one of them is forming an $I_2$-II-IV-$VI_4$ compound with one group-IV and one group-II atom by replacing two group-III atoms; another one is to replace one group-III atom and one group-I with two atoms of group-II, which formed I-II$_2$-III-VI$_4$ compound. These can also be defined by the II$_{2x}$(I-III)$_{1-x}$VI$_2$ alloy at $x$ =0.5. The I$_2$-II-IV-VI$_4$ compound especially Cu$_2$-II-IV-VI$_4$ quaternary semiconductors have been studied both experimentally and theoretically in recent research of thin film solar cell due to its optimum value of band gap (≈1.5 eV) and high-value of absorption coefficient (≈$10^4$-$10^5$ cm$^{-1}$) [3-7]. However, in literature, the studies of Cu-II$_2$-III-VI$_4$ quaternary semiconductors for example, CuCd$_2$InTe$_4$, CuTa$_2$InTe$_4$, AgZn$_2$InTe$_4$ and CuTa$_2$InTe$_4$ are synthesized in stannite or cubic structure and reported their various properties quite limited [8-17]. Using many possible atomic variations, these emerging chalcogenides groups opens new era for finding materials with suitable characteristics. A cation cross-substitution of ternary and binary compounds has been studied by Chen *et al*.[8] They evaluated the structural stability of various crystal phases and found that the stannite structure and kesterite structure have higher energy compared to the wurtzite structure after relaxation. Recently, the crystalline quaternary chaocogenides CuZn$_2$InTe$_4$ and AgZn$_2$InTe$_4$ compounds was also reported by Shi *et al* [18]. Through cross-substitutions the semiconducting chalcogenides CuZn$_2$AlSe$_4$ explored and total energy calculation provides the information of material properties, shifting of energy band and the conductivity of the material [19]. Further, a group of Cu$_2$ZnAS$_{4-x}$ (x = 0.5±0.3) and CuZn$_2$AS$_4$ (A = Al, Ga, In) nanocrystals synthesized in wurtzite-phase via hot injection method [9]. It has been found that the compounds have band gap values of range ~1.20-1.72 eV in the visible region and have high absorption values, which indicate that these compounds offer decent optical properties to be used as inexpensive and nontoxic active layers in photovoltaic applications. The structural stability, structure, and composition, band structure of CuZn$_2$AlS$_4$ has been reported earlier [10]. However, there is no systematic theoretical investigation to elucidate the structural, optoelectronic, and band structure properties of stannite and kesterite phase CuZn$_2$InS$_4$ and CuZn$_2$GaS$_4$ compounds.

In the present report, we studied the optoelectronic and structural properties of the kesterite (KS, space group I$\bar{4}$) and stannite (SS, space group I2m) type quaternary chalcopyrite compounds CuZn$_2$MS$_4$ (M =In and Ga) via full potential augmented plane wave method (FP-LAPW) through Wien2k code (Figure 1(a) and Figure 1(b)).The band structure with partial and total density of states (DOS) was evaluated from optimized structural parameters at the lowest energies. The absorption spectra, dielectric function, and refractive index were reported and discussed. In addition, we have studied the partial and total DOS of kesterite structure of $CuZn_2In_xGa_{1-x}S_4$ (where x=0.5) to explore the transition from $CuZn_2InS_4$ to $CuZn_2GaS_4$. Our systematic studies could provide the knowledge of this important group of quaternary chalcogenides and enable to compare with the related and vastly studied quaternary group.

## 2. Computational methodology

To study the phase stability from total energy calculation and the electronic structure of $CuZn_2MS_4$ (M=In and Ga) compounds, the scalar-relativistic full potential augmented plane wave method (FP-LAPW) was employed via WIEN2k code [20]. In this computational work, core and valence subsets split from the basis set located in the Muffin-tin (MT) sphere. The core state contribution comes mainly from the spherical part of the potential where the charge density is spherically symmetric.

On the other side, the scalar-relativistic approach applied for the valence states, although the full relativistic method with full relaxation counted for the core state. Here, the valence states orbital configuration are Cu ($3p^6$, $4s^2$, $3d^9$), Zn ($3d^{10}$, $4s^2$), In ($4d^{10}$, $5s^2$, $5p^1$), Ga ($3d^{10}$, $4s^2$, $4p^1$), and S ($3s^2$, $3p^4$). All these orbital expanded within a potential in spherical harmonics. The Perdew-Burke-Ernzerhof (PBE) [21] potential in generalized-gradient approximation (GGA) is applied for crystal structure optimization, optimization of atomic positions, and defect formation energy evaluation. To avoid the charge leakage inside the core state we have optimized the values of MT radii of all the atoms. The MT radii for Zn, Cu, In, Ga, and S were defined to be 2.22, 2.20, 2.30, 2.28, and 1.90 a.u. respectively. The core and valence state separation is considered in the form of cut-off energy with a value of 6.0 Ry. The self-consistent convergence of the simulations set approximately in the range of $10^{-5}$ Ry. Further, the expansion of basis was taken with $R_{MT} \times K_{MAX}$ equal to 7, where $R_{MT}$ represents the smallest atomic MT sphere radius and plane wave cut-off of k-vector defined as $K_{MAX}$. The magnitude of charge density ($G_{max}$) in Fourier expansion is equal to 12 $(Ry)^{1/2}$. In the calculation, the Brillouin zone integrations were performed with the tetrahedral method [22] with a 4×4×4 division k-point. To study the DOS of $CuZn_2In_xGa_{1-x}S_4$ (where x=0.5), we have structured a super cell from optimised $CuZn_2InS_4$ KS structure and optimised the lattice parameters of the system.

The absorption coefficient, α (ω) together with the real and imaginary part of the dielectric function [ε (ω)] also analyzed using the formula as previously reported in the literature [7].

## 3. Results and discussion

### 3.1 Total energy

The kesterite and stannite phases of CZMS are found to be more stable compared to the wurtzite phase. The total equilibrium energy of wurtzite, kesterite, and stannite crystal phases are checked. The optimized atomic position of the more stable KS and SS are tabulated in Table 1.The variation of total energy with the volume of the compound is shown in Fig.2. The equilibrium lattice constant was evaluated from the volume with the corresponding value of lowest energy. The optimized lattice parameters are evaluated as a=5.529 Å, c=11.058 Å for $CuZn_2InS_4$_KS, a=5.533 Å, c=11.066 Å for $CuZn_2InS_4$_SS, a=5.417 Å, c= 10.833 Å for $CuZn_2GaS_4$_KS, a=5.421 Å, c=10.842 Å for $CuZn_2GaS_4$_SS.

### 3.2 Density of states (DOS) and Band structure

To analyze the electronic behaviour of the quaternary compounds, the band structure and DOS of all the compounds in KS and SS structures are plotted in Fig. 3 and Fig. 4, respectively. It observed that the band gap decreases with increasing atomic no of group-III elements within the compound. As the conduction band minimum (CBM) is composed of group III-s and anion p states, the band gap is reduced from Ga to In. From the DOS plots, it observed that the valence band maximum (VBM) is coming from the Cu *3d* and anion *p* states, as plotted in Fig. 3.Generally, The KS structure is more stable and has a lesser band gap as compared with SS structure of the material. It can be seen that all the structures are direct-band semiconductors. Further, it observed that movement of Ga to In, the bonding and anti-bonding states of *p−d* orbitals overlap more. For all the compounds, it is clear from DOS plots that the VBM state leads by the antibonding *p−d* coupling between the group I (i.e. Cu) and VI (i.e. S) atoms, while CBM states are dominated by the antibonding coupling between the group III (Ga/In) *s* state and VI (i.e. S) *s* and *p* states. Therefore, very little shifting of VBM occurs and due to the remarkable downshifting CBM from In to Ga, the band-gap value decreases. The band structure in Fig. 4 is plotted for further understanding of the electronic properties. It also shows the direct band nature of CZMS, similar to the reported literature of $CuZn_2GaS_4$ [9] and $CuZn_2AlS_4$ compounds [10].

Fig. 5 displays the overall total DOS together with the partial DOS of *Cu-s, Cu-d, Zn-s, In-s, d, Ga-s, Ga-d, S-s* and *S-p* orbitals of $CuZn_2In_xGa_{1-x}S_4$ (where x=0.5) in kS structure. The calculated band gap value is 1.52 eV. The relative band gap of $CuZn_2In_xGa_{1-x}S_4$ bit decrease from CZGS because of downshifting of conduction band side. This occurs due to larger ionic radius of In compared with Ga. The trend of DOS is similar with CZGS and CZIS structure. Around the Fermi energy ($E_F$), the *d* orbital of *Cu* hybridized with S-*p* in edge of valance band. Also the admixture of *Cu-d* and the *d*-orbital of *In* and *Ga* hybridizes with *S-p* states in energy range of -6-0 eV. On the conduction band side the major contribution take place with *In-s, Ga-s* and *S-p* states.

### 3.3 Optical Properties

The dielectric function of all the structures is depicted in Fig. 6 with energy from 0 to 10 eV. We have found that dielectric functions have quite similar trends over a broader energy range. The real part of the dielectric function, $\varepsilon_1(\omega)$ can be defined by the dispersion of the incident photons of the material, $\varepsilon_1(\omega)$, whereas the quantity imaginary part of the dielectric function, $\varepsilon_2(\omega)$ measured by the absorption of energy. In Fig. 6a, the peak around 2 eV is due to the transition of S-*3p* and Zn-*4s* orbitals.

Another key parameter for defining the optical properties of the material is the refractive index. This is related to the polarizability of ions associated with the local field inside. The refractive indices evaluated for both kesterite and stannite crystal phases of $CuZn_2MS_4$ (M =In and Ga) compound from simulation (shown in Fig. 7a) are 2.58, 2.60, 2.66, and 2.67 respectively. From the figure, we found that the estimated refractive index for all the crystal phases in the visible region increases and attains the UV region with its maximum values. It further decreases with increasing the energy. Since with increasing value of refractive index, band gap of the material decreases, so this properties helps solar cell to respond more in higher wavelength region of the solar spectrum.

The full spectrum of calculating the absorption coefficient for both compounds is shown in Figure 7b. The band gap values ($E_g$) of $CuZn_2MS_4$ (M =In and Ga) compound were estimated from the Tauc's plot. In this plot, we considered the absorption coefficient (α) as $\alpha^2 \propto (h\nu - E_g)$, where $h\nu$ is the energy photon (υ is the frequency *and h* is the Planck's constant). With considering scissor correction, the calculated optical band gaps using PBE are 1.44 eV, 1.48 eV, 1.54 eV, and 1.57 eV for $CuZn_2InS_4\_KS$, $CuZn_2InS_4\_SS$, $CuZn_2GaS_4\_KS$, and $CuZn_2GaS_4\_SS$ respectively (inset figure of figure 7(b)).

### 4. Conclusions

In summary, a systematic investigation of the structural, optical, and electronic properties of $CuZn_2InS_4$ and $CuZn_2GaS_4$ in the KS and the SS phases was carried out using the density functional scalar-relativistic full potential linear augmented plane wave method. The density of states, band diagram, and optoelectronic properties, such as absorption coefficient, dielectric function, and refractive index have been reported in detail. From the total energy calculation, we found that the kesterite and the stannite crystal phases are more stable compared to the wurtzite crystal phase. In the case of heavier group-III atoms, the bonding and anti-bonding states of the compounds overlap more. Moreover, the optical spectra shifted from In to Ga towards the lower energy region.

**Acknowledgments**

A. Ghosh acknowledges the Science and Engineering Research Board (SERB), Department of Science and Technology (DST), Government of India (New Delhi). Also, A. Ghosh like to thank SR University for the necessary computational facilities to complete the work.

**Author contributions**

Anima Ghosh and R. Thangavel conceived and directed the project. Anima Ghosh did the simulation work and analyzed the data. Anima Ghosh and R. Thangavel participated in the preparation of the manuscript and commented on its content.

**Conflict of interest:** The authors declare no conflict of interest.

**Data and code availability:** Not Applicable

**Supplementary information:** Not Applicable

**Ethical approval:** Not applicable

**Figures**

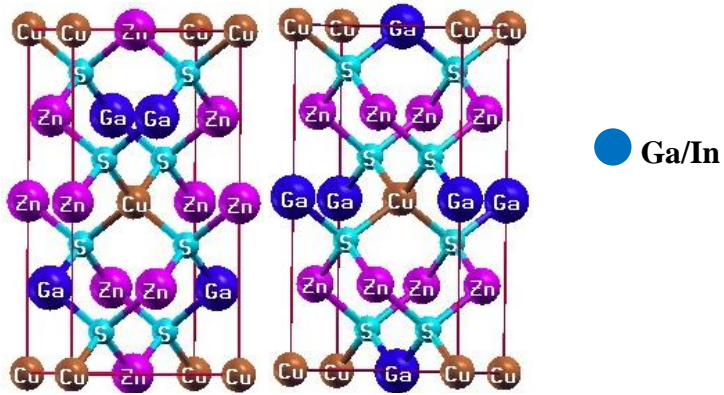

**Fig.1.** Crystal structures of KS and SS structure of $CuZn_2GaS_4$ respectively.

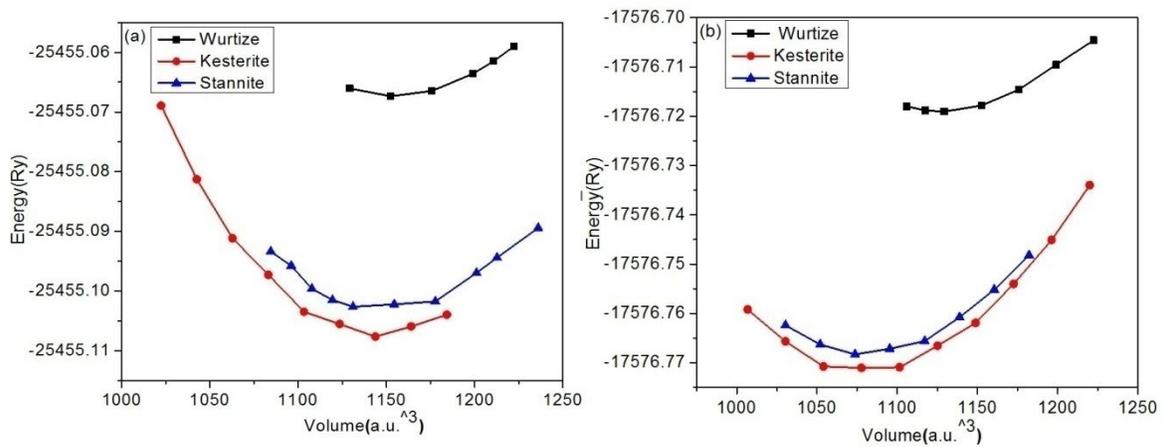

**Fig.2.**(a) and (b) represent volume vs. total energy for $CuZn_2InS_4$ and $CuZn_2GaS_4$ respectively.

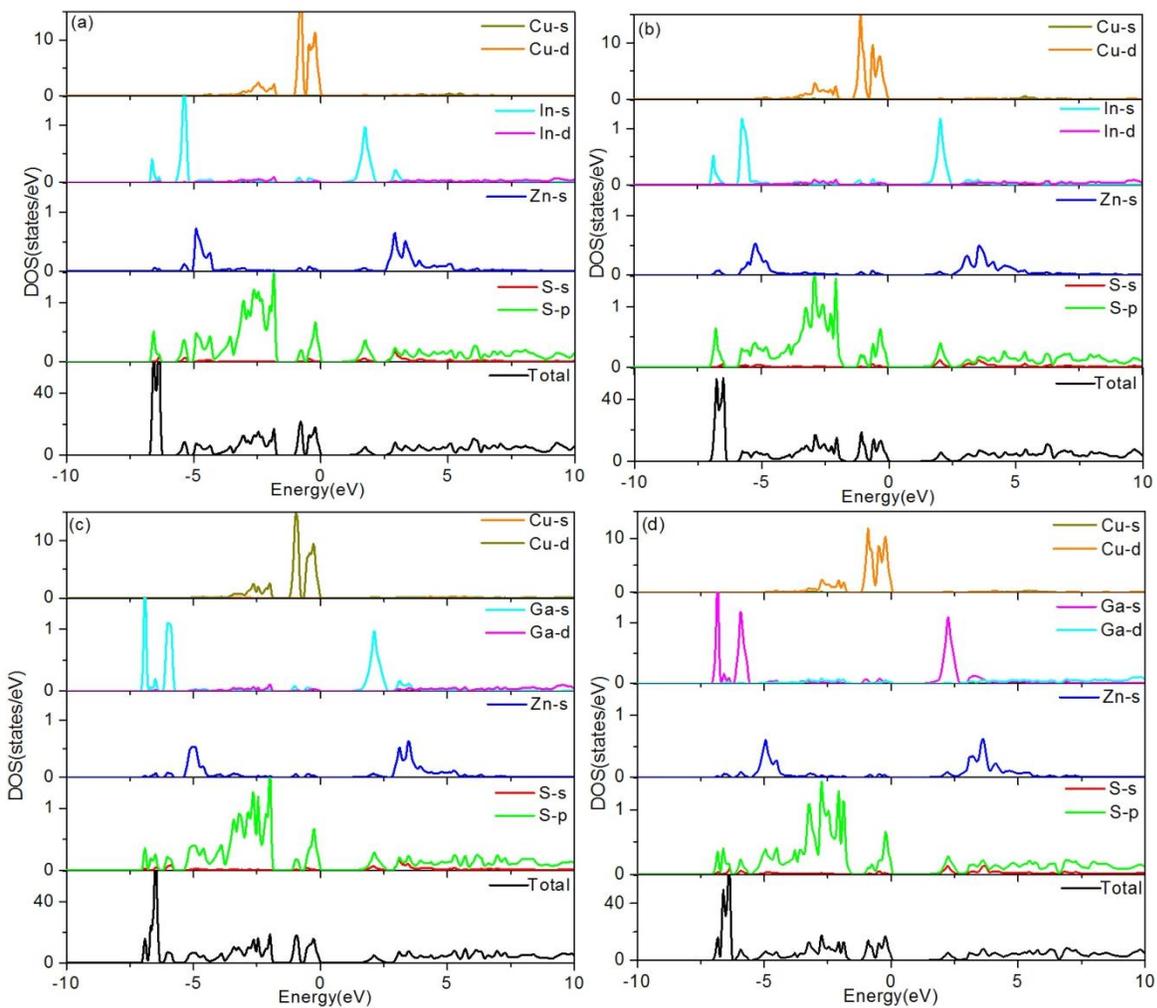

**Fig.3.** Partial density of states (DOS) and total DOS of KS-CuZn$_2$InS$_4$, SS- CuZn$_2$InS$_4$, KS-CuZn$_2$GaS$_4$ and SS-CuZn$_2$GaS$_4$.

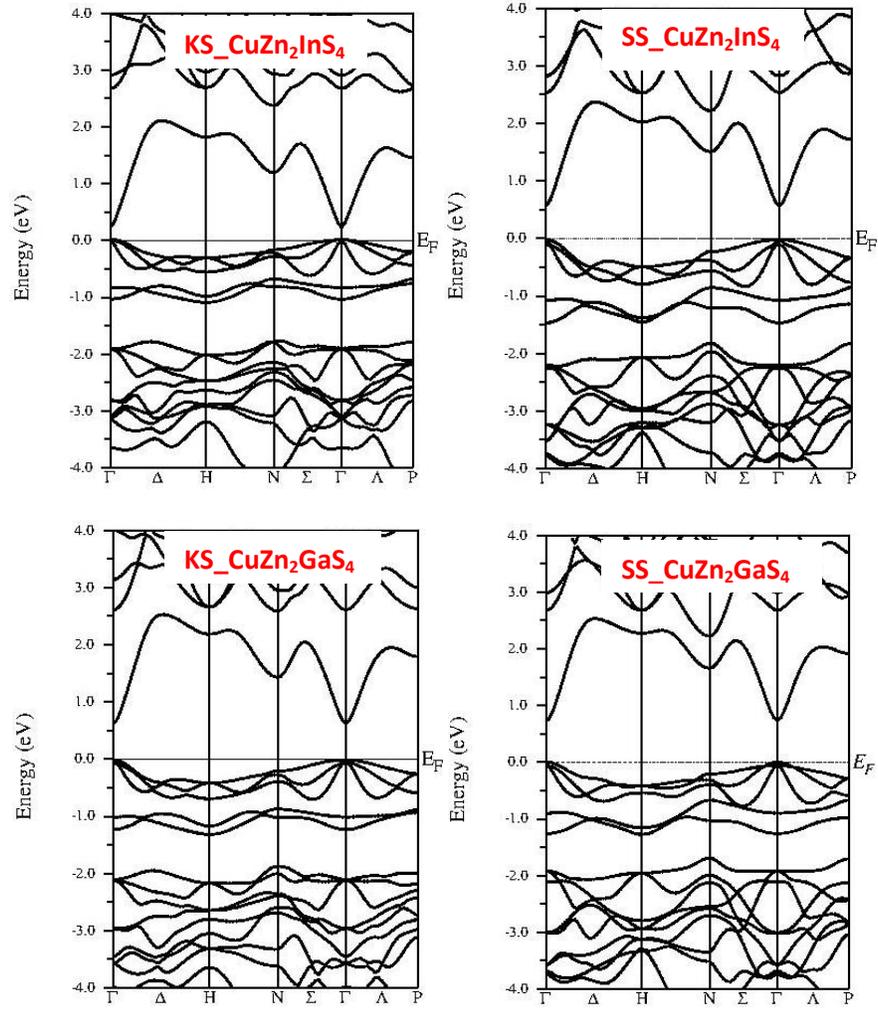

**Fig. 4.** Band structure spectrafor KS-CuZn$_2$InS$_4$, SS- CuZn$_2$InS$_4$, KS-CuZn$_2$GaS$_4$, and SS-CuZn$_2$GaS$_4$ respectively.

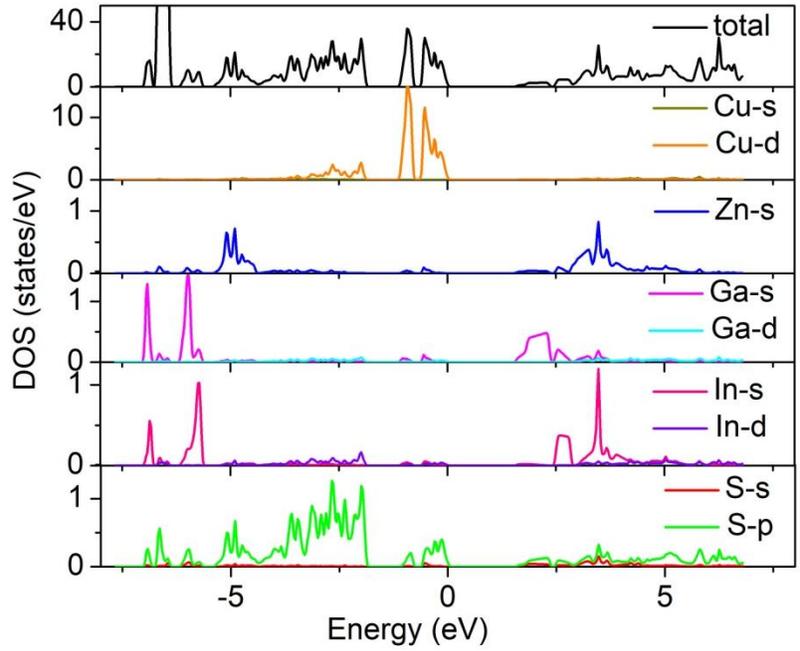

**Fig.5.** Partial density of states (DOS) and total DOS of KS-CuZn$_2$Ga$_{0.5}$In$_{0.5}$S$_4$.

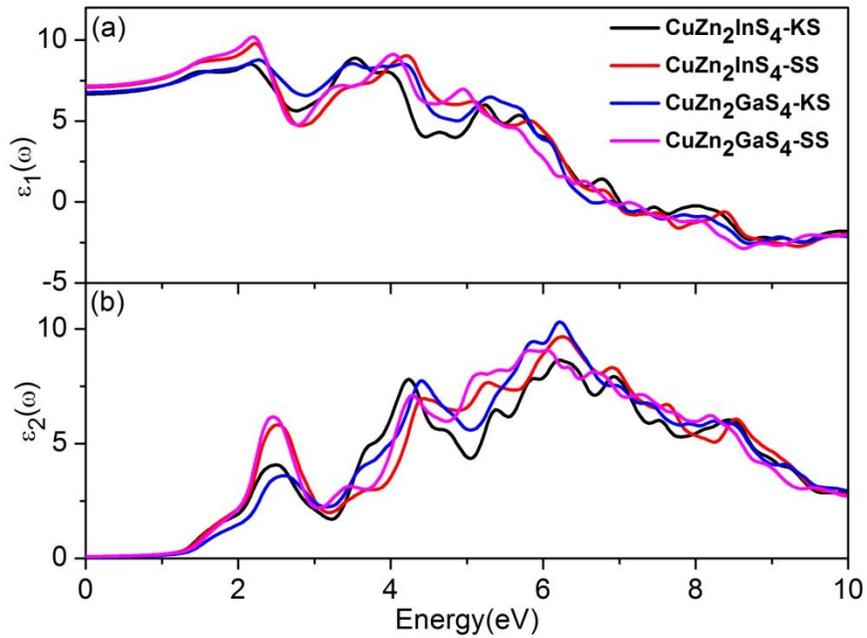

**Fig.6.** Dielectric function spectra $\varepsilon(\omega)=\varepsilon_1(\omega)+i\varepsilon_2(\omega)$ of CuZn$_2$InS$_4$, CuZn$_2$InS$_4$, CuZn$_2$GaS$_4$ and CuZn$_2$GaS$_4$ in KS and SS structures.

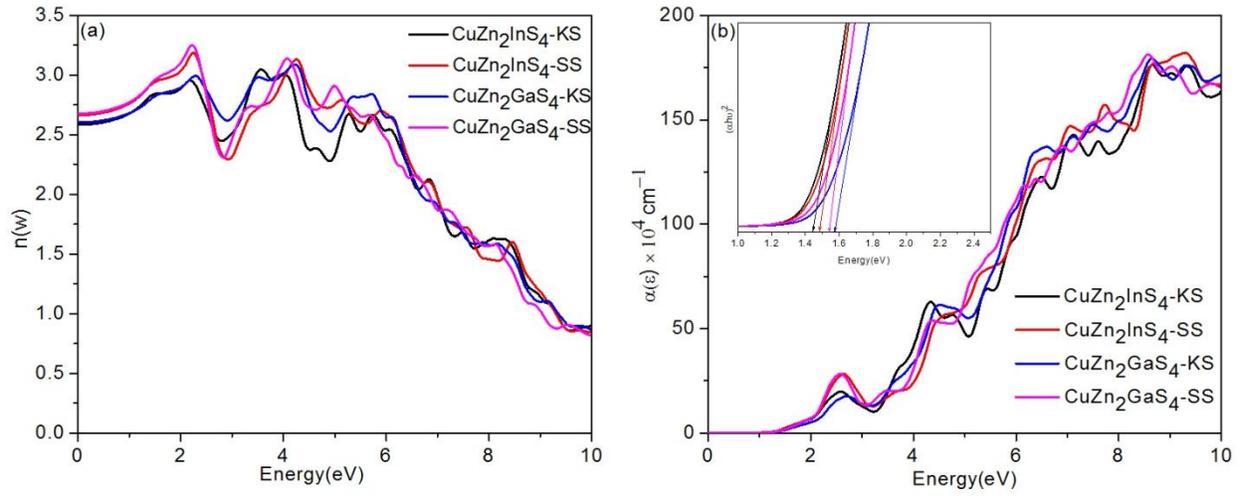

**Fig.7.** Refractive index (n) and absorption coefficient (α) spectra of CuZn$_2$InS4, CuZn$_2$InS4, CuZn$_2$GaS4, and CuZn$_2$GaS$_4$ in KS and SS structures. The inset figure within (b) shows the tauc plots of all structures respectively.